\documentclass[useAMS,usenatbib,usegraphicx]{report}
\usepackage{amsmath,amsfonts,amssymb}
\usepackage[latin1]{inputenc}
\usepackage{times} 
\usepackage{color}     
\usepackage{graphicx}
\usepackage{pdfpages}
\usepackage[normalem]{ulem} 
\usepackage{epstopdf}
\usepackage{epsfig}

\newenvironment{shortitem}
{\begin{list}{$\bullet$}{\topsep=0pt\itemsep=0pt\parsep=0pt\parskip=0pt}}
{\end{list}}

\newcommand{\Fig}[1]{Figure~\ref{#1}}
\newcommand{\Figs}[1]{Figures~\ref{#1}}

\newcommand{\Tab}[1]{Table~\ref{#1}}

\newcommand{\Float}{{\tt Real(32bit)}}
\newcommand{\Long}{{\tt Int(32bit)}}
\newcommand{\LongLong}{{\tt Int(64bit)}}
\newcommand{\Str}{{\tt String}}
\newcommand{\halos}{haloes}

\newcommand{\hdf}{HDF5}

\newcommand{\smtw}{{\sc Sussing Merger Trees} Workshop}

\newcommand{\consistenttree}{\textsc{Consistent Trees}}
\newcommand{\dtree}{\textsc{D-Trees}}
\newcommand{\mergertree}{\textsc{MergerTree}}

\newcommand{\hbt}{\textsc{HBT}}
\newcommand{\jmerger}{\textsc{JMerge}}

\newcommand{\LHaloTree}{\textsc{LHaloTree}}
\newcommand{\sublink}{\textsc{SubLink}}
\newcommand{\treemaker}{\textsc{TreeMaker}}
\newcommand{\velociraptor}{\textsc{VELOCIraptor}}
\newcommand{\ysamtm}{\textsc{ySAMtm}}

\newlength{\figwidth}
\newlength{\figtable}
\newlength{\figtripple}
\newlength{\resplot}
\title[Merger Tree Data Format]
{Sussing Merger Trees: A proposed Merger Tree data format}

\author[Sussing Merger Trees Workshop participants]
{Peter~A.~Thomas$^1$,\thanks{Email: p.a.thomas@sussex.ac.uk} 
 Julian~Onions$^2$, 
 Dylan~Tweed$^{3,4}$, 
 Andrew~J.~Benson$^5$,\newauthor
 Darren Croton$^6$,  
 Pascal~Elahi$^7$,
 Bruno~Henriques$^8$, 
 Ilian~T.~Iliev$^1$,\newauthor
 Alexander~Knebe$^{9,10}$, 
 Hanni Lux$^{2,11}$, 
 Yao-Yuan~Mao$^{12,13,14}$,
 Mark~Neyrinck$^{15}$,\newauthor
 Frazer~R.~Pearce$^2$, 
 Vicente~Rodriguez-Gomez$^{16}$, 
 Aurel~Schneider$^{17}$,
 Chaichalit~Srisawat$^1$\\
  $^1$Department of Physics \& Astronomy, University of Sussex, Brighton, BN1 9QH, UK\\
  $^2$School of Physics \& Astronomy, University of Nottingham, Nottingham, NG7 2RD, UK\\
  $^3$Center for Astronomy and Astrophysics, Department of Physics, Shanghai Jiao Tong University, Shanghai 200240, China\\
  $^4$Racah Institute of Physics, The Hebrew University, Jerusalem 91904, Israel \\ 
  $^5$Carnegie Observatories, 813 Santa Barbara Street, Pasadena, CA 91101, USA\\
  $^6$Centre for Astrophysics and Supercomputing, Swinburne University of Technology, Hawthorn, Victoria 3122, Australia\\
  $^7$Sydney Institute for Astronomy, University of Sydney, Sydney NSW 2016, Australia\\
  $^8$Max-Planck-Institut f\"ur Astrophysik, Karl-Schwarzschild Strasse 1, Garching bei M\"unchen, Germany\\
  $^9$Departamento de F\'isica Te\'orica, M\'odulo C-15, Facultad de Ciencias, Universidad Aut\'onoma de Madrid, 28049 Cantoblanco, Madrid, Spain\\
  $^{10}$Astro-UAM, UAM, Unidad Asociada CSIC\\
  $^{11}$Department of Physics, University of Oxford, Denys Wilkinson Building, Keble Road, Oxford, OX1 3RH, UK\\
  $^{12}$Kavli Institute for Particle Astrophysics and Cosmology, Stanford, CA 94309, USA\\
  $^{13}$Physics Department, Stanford University, Stanford, CA 94305, USA\\
  $^{14}$SLAC National Accelerator Laboratory, Menlo Park, CA 94025, USA\\
  $^{15}$JHU Dept of Physics \& Astronomy, 366 Bloomberg Center, 3400 N.~Charles Street, Baltimore, MD 21218, USA\\
  $^{16}$Harvard-Smithsonian Center for Astrophysics, 60 Garden Street, Cambridge MA, 02138, USA\\
  $^{17}$Institute for Computational Science, University of Zurich, Winterthurerstrasse 190, CH-8057 Zurich, Switzerland
}

\begin{document}
\date{\today}

\pagerange{\pageref{firstpage}--\pageref{lastpage}} \pubyear{2013}\volume{0000}

\maketitle

\label{firstpage}

\begin{abstract}
  We propose a common terminology for use in describing both temporal merger
  trees and spatial structure trees for dark-matter halos.  We specify a unified
  data format in \hdf\ and provide example I/O routines in C, {\sc Fortran} and
  {\sc Python}.
\end{abstract}


\section{Introduction}

The \smtw\ held in Midhurst, Sussex on 7-12 Jul 2013 brought together tree-code
builders from 10 distinct groups: 
\consistenttree\ \citep{BWW13};
\dtree\ \citep{JHC14};
\hbt\ \citep{HJW12};
\jmerger\ (Julian Onions, unpublished);
\LHaloTree\ \citep{SWJ05};
\mergertree\ \citep{KLK10a};
\sublink\ \citep{RGV15};
\treemaker\ \citep{HDN03,TDB09};
\velociraptor\ (Pascal Elahi, unpublished);
\ysamtm\ (Intae Jung etal, in preparation).
Unpublished algorithms are briefly outlined in \citet{SMT13a}.

During that meeting, it was agreed that a common structure
was required for recording merger trees, and that it should include both spatial
and temporal information.  This paper summarises the results of that discussion
and sets out the agreed data format.  It also presents an HDF5 implementation of
that format and describes some example I/O routines that are available to
download.

\section{The anatomy of a merger tree}

We use and extend the terminology defined in \citet{SMT13a}.  For ease of
reference, we repeat that here with appropriate extensions.

We start with the basic definition of a halo
\begin{shortitem}
\item A {\bf halo} is a dark-matter condensation as returned by
  a halo-finder.
\end{shortitem}
This is a deliberately vague statement as we do not want to be too
prescriptive.  Halos will usually be identified in the output of a simulation at
a fixed time, or snapshot.  Halos may be identified at a variety of different
overdensities and may be nested within one another (however, we do not allow
halos to partially overlap).  As the simulation progresses in time, it is
natural to associate halos at one snapshot with those in subsequent ones, though
that association is not always obvious and its investigation was the main
purpose of the \smtw.

\subsection{Spatial structure}

\begin{figure}
  \centering
  \includegraphics[angle=0,width=1\linewidth]{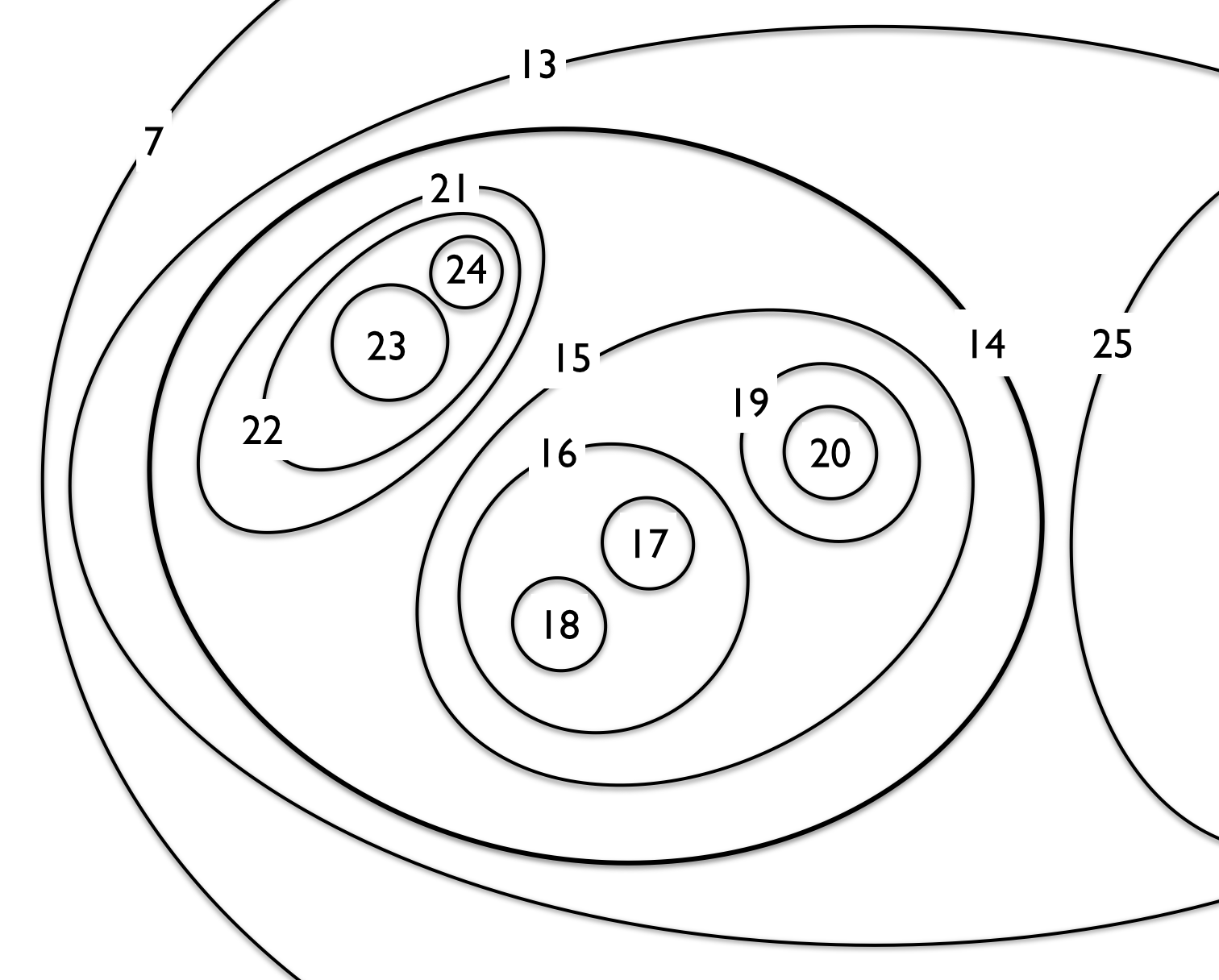}
  \caption{An example of a nested group of halos.  The numbering corresponds to
    that in the spatial structure tree representation shown in
    \Fig{fig:spatial}.}
\label{fig:halos}
\end{figure}

\begin{figure}
  \centering
  \includegraphics[width=1\linewidth]{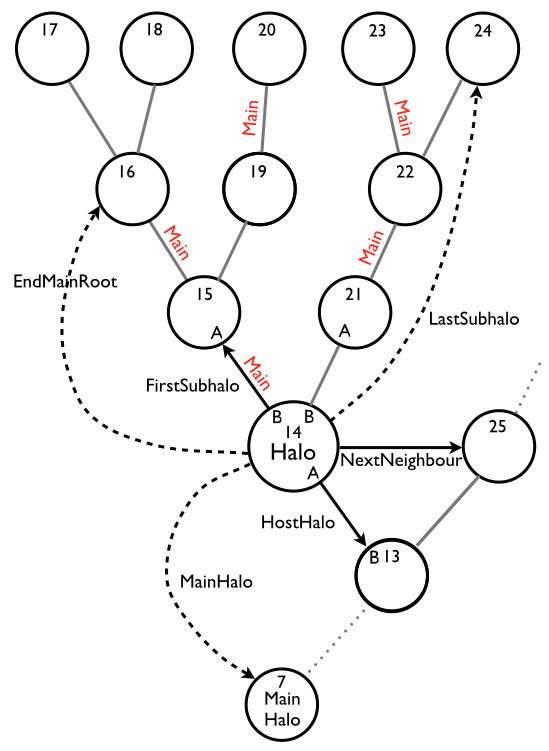}
  \caption{A representation of a spatial structure tree corresponding to the
    halos shown in \Fig{fig:halos}.  The density (size) of \halos\ decreases
    (increases) as one moves from top to bottom on the plot.  The grey lines
    show the links of the tree.  The A and B labels indicate the ordered pairs
    that are associted with Halo~14.  For that halo, the solid, black arrows
    show required links in the data structure; the designation of a FirstSubhalo
    as Main, or otherwise, and the links shown by the dashed, black arrows are
    optional.  Halos~15-24 are sub\halos\ of Halo~14; Halos~15 \& 21 are direct
    sub\halos. }
\label{fig:spatial}
\end{figure}

An example of a complex set of nested \halos\ is shown in \Fig{fig:halos}.  The
numbering of the halos is the same as that in \Fig{fig:spatial} that shows a
representation in the form a tree showing the spatial nesting of halos.

During the \smtw\ it became apparent that the current output structure of halo
finders can give difficulties in tree-building.  It is common practice to choose
one of the \halos\ in a group and designate it as the 'main halo' (also know as
the 'background halo', e.g.~\citealt{SWT01}).  This is fine for undisturbed
groups that have a relatively smooth structure with an obvious central halo.
However, it does not make sense for the coming together of two (or more) roughly
equally-sized \halos\ for which neither is at the dynamical centre of the group.
Instead we propose that all sub\halos\ be recorded separately, together with an
enclosing host halo.  Optionally, one of the sub\halos\ can then be designated a
main subhalo of the host.  The main halo is then an entity in its own right,
rather than being associated with any particular subhalo.

We propose the following nomenclature:
\begin{shortitem}
\item A {\bf spatial structure graph} is a set of ordered halo pairs,
  $(H_A,H_B)$, where $H_A$ is nested within $H_B$.  It is the purpose
  of the halo-finding codes to produce a graph that best represents
  the spatial structure of halos.  The halo definition may occur at
  discrete steps in overdensity, but this is not a requirement.
\item Recursively, $H_A$ itself and sub\halos\ of $H_A$ are {\bf
  sub\halos}\ of $H_B$.  Where it is necessary to distinguish $H_A$
  from sub\halos\ that are more deeply nested, the term
  {\bf direct subhalo} should be used.
\item Recursively, $H_B$ itself and hosts of $H_B$ are {\bf
  hosts} of $H_A$.  Where it is necessary to distinguish $H_B$
  from larger hosts, the term {\bf direct
    host} should be used.
\item Optionally, one direct subhalo may be designated the {\bf main
  subhalo} to indicate that it is at the dynamical centre of the halo.
\item The longest continuous sequence of main sub\halos\ extending
  (to higher overdensity) from a given halo is known as the {\bf main root}.
\item A halo that has no host halo is known as a {\bf main halo}.
\item A {\bf spatial structure tree} is a spatial structure graph in which
  there is precisely one direct host for every halo, except for a
  single main halo.
\end{shortitem}
In \Fig{fig:spatial}, halos are represented as circles with IDs as
indicated by the number contained within them.  The density (size) of
halos decreases (increases) from top to bottom on the plot.  

Consider Halo~14.  It has Subhalos~15-24 of which 15 \& 21 are direct subhalos.
Halo~15 has been deemed a main subhalo of 14, and Halo~16 a main subhalo of 15,
but (for the purposes of illustration) Halo 17 is not at the dynamical centre of
halo 16 and so is not designated a main subhalo.  Then Halos~14-16 are the main
root extending back from Halo~14.\footnote{We use the term root here to
  distinguish the spatial tree from the temporal one shown in \Fig{fig:temporal}
  where the equivalent structure is commonly referred to as the main branch.}
In a structure tree, each halo has a single direct host, in this case Halo~13.
The outermost enclosing main halo (the main halo) is here Halo~7.

Although not shown in \Fig{fig:spatial}, it is also useful to have
another link, StartMainRootID that shows the largest enclosing halo of
which the halo in question lies on the main root.  In the figure,
Halo~14 is not a main subhalo of 13, and so Halos 14, 15 \& 16 will
all have Halo 14 as their StartMainRoot.  This makes it easy to decide
whether a halo should be considered to be the main component of a
larger structure, and if so, which one.

\subsection{Temporal structure}

\begin{figure}
  \centering
  \includegraphics[width=1\linewidth]{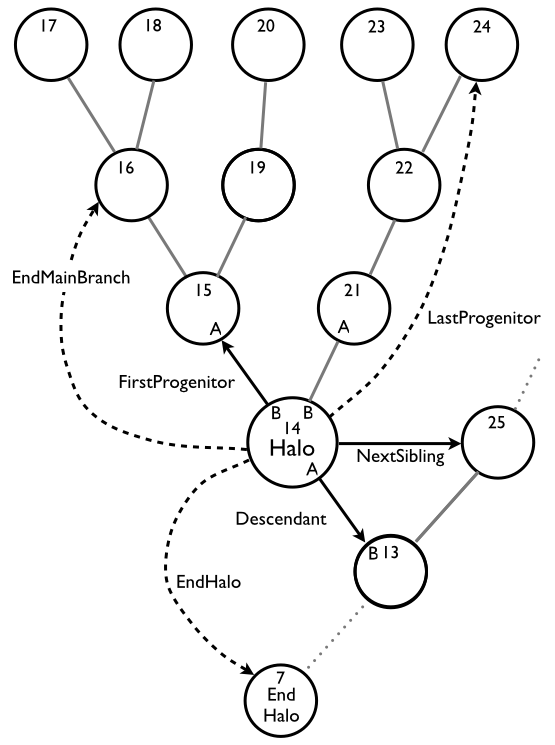}
  \caption{A representation of a temporal merger tree.  The lookback time of
    \halos\ decreases as one moves from top to bottom on the plot.  The grey
    lines show the links of the tree.  The A and B labels indicate the ordered
    pairs that are associted with Halo~14.  For that halo, the solid, black
    arrows show required links in the data structure; the links shown by the
    dashed, black arrows are optional.  Halos~15-24 are progenitors of Halo~14;
    Halos~15 \& 21 are direct progenitors.}
\label{fig:temporal}
\end{figure}

The term merger tree was first applied to temporal evolution.  As
shown in \Fig{fig:temporal}, the structure of the temporal merger tree
is identical to that of the spatial structure tree, but the nomenclature
has been changed to distinguish the two.  To maintain flexibility, we
have modified the usual definition to make the assignment of a main
progenitor optional.  To complete the analogy we designate the
\halos\ in the final snapshot (i.e.~the end of the simulation) as end
halos.

\begin{shortitem}
\item A {\bf temporal merger graph} is a set of ordered halo pairs,
  $(H_A,H_B)$, where $H_A$ is older than $H_B$.  It is the purpose of
  the merger-tree codes to produce a graph that best represents the growth of
  structure over cosmic time.  $H_A$ and $H_B$ are usually taken from
  adjacent snapshots, but this is not a requirement as there are
  occasions where \halos\ lose their identity and then reappear at a
  later time.
\item Recursively, $H_A$ itself and progenitors of $H_A$ are {\bf
  progenitors} of $H_B$.  Where it is necessary to distinguish $H_A$
  from earlier progenitors, we will use the term {\bf direct
    progenitor}.
\item Recursively, $H_B$ itself and descendants of $H_B$ are {\bf
  descendants} of $H_A$.  Where it is necessary to distinguish $H_B$
  from later descendants, we will use the term {\bf direct
    descendant}.
\item Optionally, one of the direct progenitors may be labelled the
  {\bf main progenitor} -- this will usually be the most massive,
  but other choices are permitted. (Note: this deviates from the
  definition in \citet{SMT13a} where the selection of a main progenitor
  was compulsory.)
\item The longest continuous sequence of main progenitors extending
  back in time from a given halo is known as the {\bf main branch}.
\item A halo that has no descendants is known as an {\bf end halo}.
\item A {\bf temporal merger tree} is a temporal merger graph in which there is
  precisely one direct descendant for every halo, except for a single end halo.
\end{shortitem}

\subsection{Combined spatial and temporal trees and depth-first ordering}
\label{sec:combined}

Although our standard does not prescribe it, the trees shown in
\Figs{fig:spatial} \& \ref{fig:temporal} are {\bf depth-first ordered} which
means that the subhalos of a given halo are all those with indices between
FirstSubhalo and LastSubhalo.  Similarly, the progenitors are those halos with
indices between FirstProgenitor and LastProgenitor.

\begin{figure}
  \centering
  \includegraphics[width=1\linewidth]{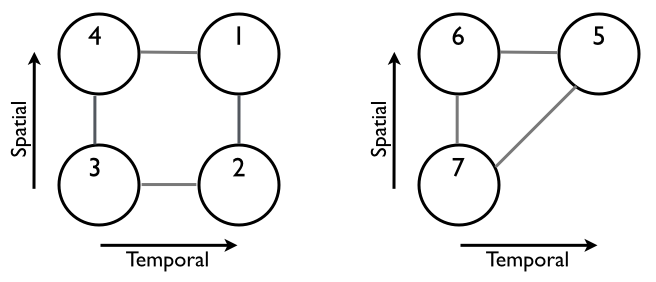}
  \caption{Simple examples of combined structural and temporal merger trees.  In
    each case, overdensity decreases upwards (i.e. enclosing structures) and
    time towards the right.  The left-hand example shows that there is no unique
    way to traverse the combined tree to reach a particular progenitor subhalo.
    The right-hand example shows that temporal links may connect a halo to
    another of differing overdensity (whereas spatial links always connect
    halos at the same time).}
\label{fig:combined}
\end{figure}

Unfortunately, there is no straightforward way to order a combined spatial and
temporal tree.  Consider the left-hand example in \Fig{fig:combined}.  Halo~3
can be reached from Halo~1 in two different ways: it is a subhalo of Halo~4
which is a progenitor of Halo~1; and it is a progenitor of Halo~2 which is a
subhalo of Halo~1.  It would seem, therefore, that we have a choice of making
either time or space the inner loop in any tree traversal.  As we describe
below, however, only the former will allow us to uniquely traverse the tree
starting from end, main halos (i.e.~the outermost enclosing halos at the
endpoint of the simulation) and visiting each halo precisely once.

As one might conceivably want to examine the structure of a halo without knowing
its merger history, it would seem natural to preserve depth-first ordering for
the former rather than the latter (i.e.~make space the inner loop in any
tree-search and time the outer one).  Unfortunately, that does not work because
the progenitors of a main halo can have a variety of overdensities and may
contain halos that are at different levels in the same spatial tree.   Consider
starting a tree traversal from Halo~5 in the right-hand example of
\Fig{fig:combined}.  Such a search would find Halo~6 as a progenitor of Halo~5
and then Halo~7 as a subhalo of Halo~6.  But it would also find Halo~7 as a
progenitor of Halo~5 in its own right and so woud appear twice in the tree search.

On the other hand, spatial trees contain halos that are all built from the same
snapshot of a simulation and so are coincident in time.  So if we search first
for subhalos in the final snapshot and then follow all those subhalos back in
time, then we will visit each halo precisely once.  For that reason, we choose
to make space the outer loop and time the inner one within which depth-first
ordering is preserved.  An example tree-search algorithm (in {\sc Python}) is
given in the Appendix.

\section{Data format}

\subsection{Halo indices and pointers}

The halos that are returned by a halo finder do not normally have consecutive
indices (i.e.~with no gaps in the list).  However, to allow for efficient data
retrieval, the pointer indices that represent the links in the figures need to
run in a continuous sequence from HaloIndexOffset to HaloIndexOffset+Nhalo-1,
where Nhalo is the total number of halos.\footnote{The use of HaloIndexOffset
  allows for an offset in case that the indexing does not start from zero
  (e.g.~the use of {\sc Fortran}-style indexing, or when working with a subset
  of a larger data-set).}  We therefore make this a requirement but also store
the original HaloID to allow cross-reference back to the halo catalogue.

To avoid repetition, we describe the structure of the data format here only for
the structure tree; that of the merger tree is exactly analogous.  To store a
basic tree in which there is no main subhalo designation, then all that would be
required would be a pointer to the direct host of each halo -- this is the
HostHaloIndex.  However, in order to allow efficient traversal of the tree, we
also need an efficient way to locate subhalos of a given halo.  That is provided
by two pointers: the FirstSubhaloIndex points to the main subhalo, if that
exists, or to one of the other direct sub\halos\ if not (typically that with the
lowest halo index).  Then each of those sub\halos\ is also given a
NextNeighbourIndex pointer to another of the direct sub\halos\ (typically the
one with the next highest halo index) such that eventually all the direct
subhalos have been visited.  If any of the halos described above does not exist,
then the corresponding pointer is set to a null value (e.g.~$-1$ for C/{\sc
  Python} or 0 for {\sc Fortran}).

If the halo indexing is done in depth-first order, as shown in the figures, then
the (optional) inclusion of pointers EndMainRootIndex to indicate the end of the
main root, and LastSubhaloIndex to identify the subhalo with the largest index,
allows for efficient identification of contiguous blocks of useful data.

The inclusion of MainSubhaloFlag for identification of the first subhalo as a
main subhalo (or not) is also optional.  For trees in which the first subhalo is
always a main subhalo then this could be omitted.  Likewise, it is common
practice to always require a main progenitor but the (optional) inclusion of a
MainProgenitorFlag allows flexibility not to do so, for example in the case of
equal-sized mergers in which there is no obvious dominant progenitor on the
previous snapshot.

\subsection{\hdf\ data format}

The use of \hdf\ allows for efficient data compression and avoids having to
worry about the endianness of the data.  \Tab{tab:hdf} outlines an
\hdf-compatible data format to record both spatial and temporal merger trees.
Required entries are entered in plain text and optional ones in italics.  The
required entries specify minimal information to associate halos with particular
snapshots and density levels and to allow efficient traversal of the tree.  We
also include the halo mass as that is a fundamental porperty that will be
required in almost any application.

To increase flexibility, we allow the entries in the MergerTree group to be
stored either in a table (one row per halo) or as a set of arrays (one array per
property; Nhalo entries per array).  Which of these is more efficient in terms
of data I/O will depend upon the number of halos and the number of properties
that are stored.

Note that the pointers refer to the position of the corresponding halo in this
array of halo structures, not to the original halo ID in the halo catalogue
(this latter OriginalHaloID may optionally be provided separately to allow
cross-reference).  The pointers thus have values between HaloIndexOffset and
HaloIndexOffset$+$Nhalo$-$1, inclusive.  A null pointer can be taken as any
number that lies outside this range, although the values $-1$ and 0 are the most
common (for {\sc C/Python} and {\sc Fortran}, respectively).

We give two options for storing the physical properties of halos -- in practice
users may want to use a mixture of the two.  Firstly, for ease of lookup,
properties may be specified in the MergerTree group; alternatively, they may be
stored by snapshot number in groups Snapshots/Snapshot\#.  In order to look up
the latter a HaloIndex pointer would be needed to identify the appropriate entry
in the table.

In general a data-set may contain more than one tree and in that case
it may be useful to know the indices of the main (end) halos.  Our
suggested data structure avoids storing that information as it would
be contained in an array of variable length, but we note that it can
easily be recovered by searching for \halos\ for which the HostHaloIndex
(DescendantIndex) is the null pointer.

Note that arrays are datasets with a predefined \hdf\ datatype.  Arrays and
attributes can be manipulated using the high-level H5LT routines.  Tables have
compound datatypes and are manipulated using the high-level H5TB routines.

\begin{table*}
  \caption{An interpretation of the data format in \hdf.}
  \label{tab:hdf}
\begin{tabular}{lllll}
\hline
Group& Object name& Object type& Notes\\
\hline
{\bf /}\\
& Version& \Long\ attribute& Version number\\
& Subversion& \Long\ attribute& Subversion number\\
& Title & \Str\ attribute & Title of data-set\\
& Description & \Str\ attribute & Description of data-set\\
& BoxsizeMpc & \Float\ attribute & Comoving box-size in Mpc ({\bf not} $h^{-1}$Mpc)\\
& OmegaBaryon & \Float\ attribute & Density parameter of baryons at $z=0$\\
& OmegaCDM  & \Float\ attribute & Density parameter of CDM
 (excluding baryons) at $z=0$\\
& OmegaLambda & \Float\ attribute & Density parameter of Cosmological
Const. ($w=-1$) at $z=0$\\
& H100 & \Float\ attribute & Hubble parameter / 100\,km\,s$^{-1}$\,Mpc$^{-1}$\\
& Sigma8 & \Float\ attribute & RMS linear overdensity at $z=0$ in spheres
 of radius 8\,$h^{-1}$\,Mpc\\
& \ldots& {\it Attribute}& {\it Optional additional attributes}\\
& Snapshots & Group& Snapshot info/data\\
& MergerTree& Group& Merger tree info/data\\ 
\hline
/Snapshots\\
& NSnap& \Long\ attribute & Number of snapshots\\
& Snap& Table& Table of snapshot properties (a, t, \ldots)\\
& SnapProp& Table& Table of snapshot property names/descriptions/units\\
& {\it Snapshot1}& {\it Group} & {\it Snapshot particle and/or halo data}\\
& \ldots & \ldots& \ldots\\
& {\it SnapshotNsnap}& {\it Group} & {\it Snapshot particle and/or halo data}\\
\hline
\multicolumn{2}{l}{\it /Snapshots/Snapshot\#}\\
& NSnapHalo& \Long\ attribute & Number of halos in this snapshot\\
& SnapHalo& Table& Table of halo properties (Mass, Pos, Vel, AngMom, \ldots)\\
& SnapHaloProp& Table& Table of halo property names/descriptions/units\\
\hline
/MergerTree&\\
& NHalo& \Long$|$\LongLong\ attribute& Total number of halos\\
& HaloIndexOffset& \Long$|$\LongLong\ attrribute& Pointer index value of first halo in the Arrays/Tables\\
& TableFlag& \Long\ attribute& 0$|$1 if halos stored in Arrays$|$Table\\
& \ldots& \ldots& Optional additional attributes\\
\multicolumn{2}{l}{\it Either (table format):}\\
& Halo& Table& Table of halo properties (Snapshot, Density, Mass,
FirstSubhaloIndex, \ldots)\\
& HaloProp& Table& Table of halo property names/descriptions/units\\
\multicolumn{2}{l}{\it or (arrays of size NHalo$^\dagger$):}\\
& Snapshot & \Long\ & Snapshot number\\
& Density & \Float\ & Over-density of halo (units to convert to
critical density)\\
& Mass & \Float\ & Some measure of halo mass\\
& FirstSubhaloIndex & \Long$|$\LongLong\ & Pointer to FirstSubhalo\\
& NeighbourIndex & \Long$|$\LongLong\ & Pointer to Neighbour\\
& HostHaloIndex & \Long$|$\LongLong\ & Pointer to HostHalo\\
& FirstProgenitorIndex & \Long$|$\LongLong\ & Pointer to FirstProgenitor\\
& NextSiblingIndex & \Long$|$\LongLong\ & Pointer to NextSibling\\
& DescendantIndex & \Long$|$\LongLong\ & Pointer to Descendant\\
& {\it EndMainRootIndex} & \Long$|$\LongLong\ & {\it Pointer to halo at the end
  of the main root}\\
& {\it EndMainBranchIndex} & \Long$|$\LongLong\ & {\it Pointer to halo at the end
  of the main branch}\\
& {\it LastSubhaloIndex} & \Long$|$\LongLong\ & {\it Pointer to LastSubhalo}\\
& {\it LastProgenitorIndex} & \Long$|$\LongLong\ & {\it Pointer to LastProgenitor}\\
& {\it MainSubhaloFlag}& \Long\ & {\it 0$|$1 if FirstSubhalo is a main subhalo}\\
& {\it MainProgenitorFlag}& \Long\ & {\it 0$|$1 if FirstProgenitor is a main progenitor}\\
& {\it OriginalHaloID}& \Long$|$\LongLong\ & {\it Halo ID in original catalogue}\\
& {\it HaloIndex}& \Long$|$\LongLong\ & {\it Location in
  Snapshot Halo table}\\
& {\it Pos}& \Float\ {\it array[3]$^\ddagger$} & {\it Comoving position of
  centre of mass}\\
& {\it Vel}& \Float\ {\it array[3]$^\ddagger$} & {\it Peculiar Velocity of
  centre of mass}\\
& {\it AngMom}& \Float\ {\it array[3]$^\ddagger$} & {\it Angular momentum of halo}\\
& \ldots& \ldots& {\it Other halo properties}\\
\hline
\multicolumn{4}{l}{{}$^\dagger$Each array to come with string Description and Units attributes.}\\
\multicolumn{4}{l}{{}$^\ddagger$i.e.~arrays of size [NHalo,3] or [3,NHalo]
  dependent upon the ordering of the data (C-like or {\sc Fortran}-like, respectively).}\\
\multicolumn{4}{l}{{}Italic text indicates optional entries.}
\end{tabular}

\end{table*}

\subsection{Example I/O routines}

One of the main advantages of \hdf\ and similar formats is that the
structure of the file and the nature of its contents can be discerned using
utility programs such as {\tt hdfview}, {\tt h5ls} and {\tt h5dump}.

To encourage use of the data format described in this paper, we provide example
programs to read and write merger tree data files in {\sc C}, {\sc Fortran-95}
and {\sc Python}.  For those new to \hdf\ we strongly recommend first working
with {\sc Python} as the {\tt h5py} module greatly simplifies the interface with
the underlying \hdf\ libraries.  In applications handling large data-sets,
however, the greater efficiency of {\sc C} or {\sc Fortran} will be required.

We provide some basic examples of merger tree I/O in {\sc Python} in the
Appendix.  These routines, together with their equivalents in {\sc C}
\& {\sc Fortran} may be downloaded from {\tt
  https://bitbucket.org/ProfPAThomas/mergertree/}.  These routines
are very much work in progress and we encourage anyone with the interest and
time to improve and extend them.

\section*{Acknowledgements} \label{sec:Acknowledgements}

The {\sc Sussing Merger Trees} Workshop was supported by the European
Commission's Framework Programme 7, through the Marie Curie Initial
Training Network CosmoComp (PITN-GA-2009-238356).  This also provided
fellowship support for AS.

KD acknowledges the support by the DFG Cluster of Excellence "Origin
and Structure of the Universe". 
 
PJE is supported by the SSimPL programme and the Sydney Institute for
Astronomy(SIfA), and through the ARC via DP130100117.

AK is supported by the {\it Ministerio de Econom\'ia y Competitividad} (MINECO)
in Spain through grant AYA2012-31101 as well as the Consolider-Ingenio 2010
Programme of the {\it Spanish Ministerio de Ciencia e Innovaci\'on} (MICINN)
under grant MultiDark CSD2009-00064. He also acknowledges support from the {\it
  Australian Research Council} (ARC) grants DP130100117 and DP140100198. He
further thanks Primal Scream for velocity girl.

HL acknowledges a fellowship from the European Commissions Framework Programme
7, through the Marie Curie Initial Training Network CosmoComp
(PITN-GA-2009-238356).

CS is supported by The Development and Promotion of Science and Technology 
Talents Project (DPST), Thailand.

PAT acknowledges support from the Science and Technology Facilities
Council (grant number ST/L000652/1).

The authors contributed in the following ways to this paper: PAT initiated the
discussion and wrote the paper.  PAT \& JO wrote the provided computer code.
The other authors participated in the discussion and design of the data format,
in particular JO \& DT.  All authors helped to proof-read the paper.

\bibliography{mn-jour,tree}
\bibliographystyle{mn2e} \label{sec:Bibliography}



\section*{Supplementary material}

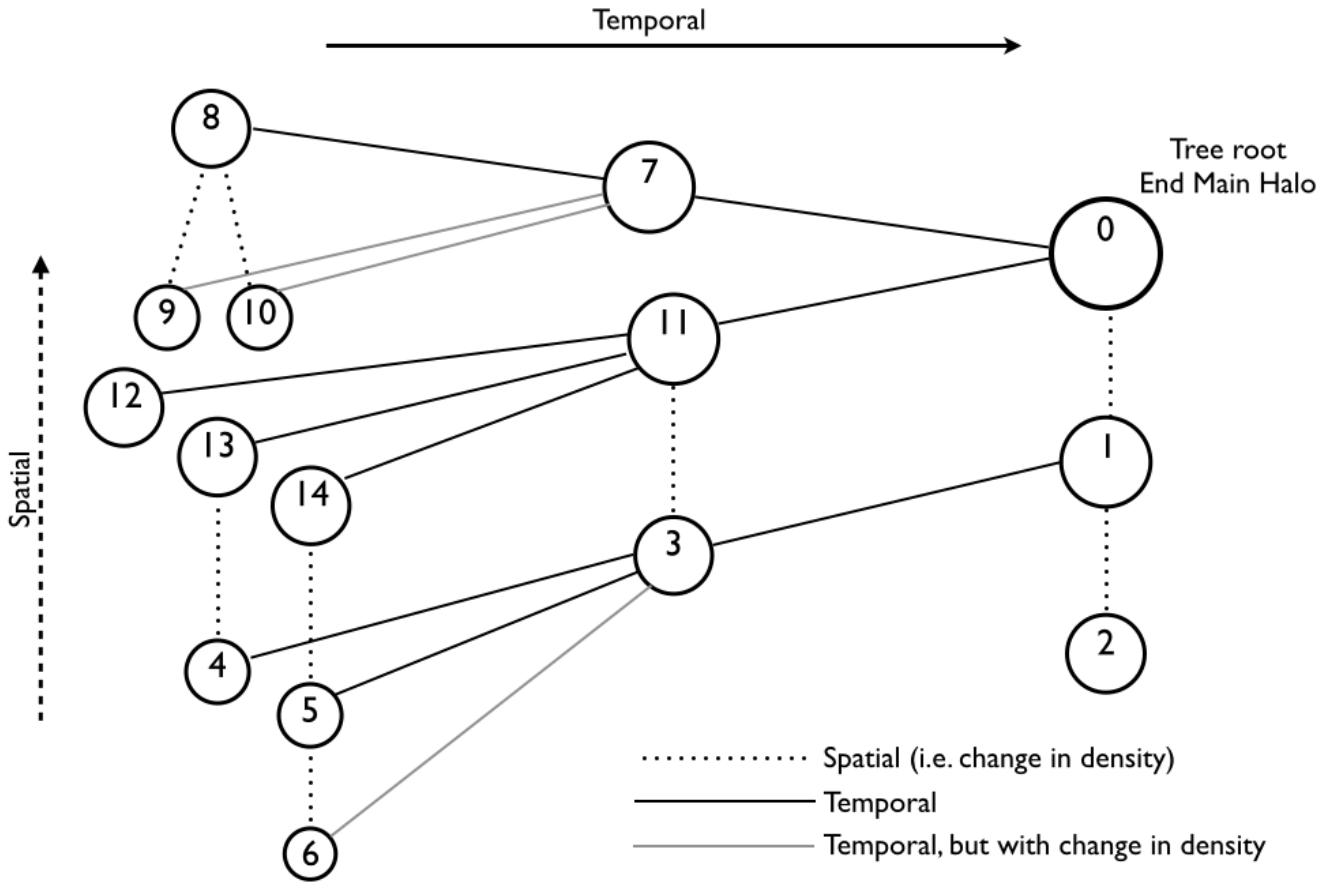

**Figure A1.** A test merger tree that is used for the examples in this appendix.

## APPENDIX A: EXAMPLE PYTHON CODES

In this appendix we present some example PYTHON codes to search merger trees and to read/write them in HDF5 format. These routines can all be downloaded from https://bitbucket.org/ProfPAThomas/mergertree/.

The examples use the tree shown in Figure A1 as recorded in HDF5 format in test.hdf5.

### A1   To read a tree written in the default format

We first present an example of how to read in data from an HDF5 file and to dump some properties to the screen. To simply the script, this example makes use of mt.py that defines a MergerTree class and some helper functions, but this is not really necessary and could easily be omitted.

```
"""readtree.py
   Example of how to read in a merger tree in HDF5 format
   Usage: python readtree.py <tree.hdf>
"""

#-----------------------------------------------------------------------------
# Imports

# The following imports the python3 print function into python2;
# It seems to be quietly ignored in python3
from __future__ import print_function

import sys
```

```python
# The merger tree module contains data structures, methods and global variables
import mt

#-------------------------------------------------------------------------------
# Open file
if len(sys.argv) != 2: sys.exit("Usage: " + sys.argv[0] + " filename.hd5")
mtdata=mt.MergerTree(sys.argv[1], "r")
for (name,value) in mtdata.attrs('/'): print(name,value)

# Read in snap table
for (name,value) in mtdata.attrs("Snapshots"): print(name,value)
snap=mtdata.read_snaptable()
print('Snap =')
print('{0:>8} {1:>6} {2:>6} {3:>6}'.format(*snap.dtype.names))
for row in range(len(snap)): print('{0:>8} {1:>6.3f} {2:>6.3f} {3:>6.3f}'.format(*snap[row]))
del snap

# Read merger tree from HDF file
for (name,value) in mtdata.attrs("MergerTree"): print(name,value)
tree=mtdata.read_mergertree()
print('Tree =')
for dset in tree:
    print(dset,end=' ')
print()
for ntree in range(min(10,len(tree['HaloID']))):
    for prop in tree:
        print(tree[prop][ntree],end=' ')
    print()
del tree

#Close file
mtdata.close()

#-------------------------------------------------------------------------------
```

### A2 Merger tree class and associated methods

Here is file mt.py containing the MergerTree class used in the above example.

```python
"""mt.py
   Merger tree class to simplfy some i/o functions for merger trees.
   It's not entirely obvious that a separate class is needed
   - could just put these calls directly into the main program.
"""

#-------------------------------------------------------------------------------
import h5py

#-------------------------------------------------------------------------------
class MergerTree:

    def __init__(self, file=None, mode='r'):
        """Initialise the class, opening the file with the mode given"""
        if file:
            self.open(file, mode)
        else:
```

```
            print('mt usage: mt(file, mode=mode)')

    # Return set of attributes associated with object
    def attrs(self, object):
        return self.fid[object].attrs.items()

    # Close the HDF5 file
    def close(self):
        self.fid.close()

    # Open the HDF5 file
    def open(self, file, mode):
        self.fid = h5py.File(file, mode)

    # Return a handle to the MergerTree group
    def read_mergertree(self):
        return self.fid["MergerTree"]

    # Return a handle to the Snapshot table
    def read_snaptable(self):
        return self.fid["Snapshots/Snap"]

    # Return a handle to the SnapProp table
    def read_snapprop(self):
        return self.fid["Snapshots/SnapProp"]

#--------------------------------------------------------------------------------
```

### A3  Tree-searching

The following example tracetree.py reads in an HDF5 file then shows how to loop over (or 'trace') a tree using a spatial and/or temporal search. It makes use of a Tree class defined in tree.py

```
"""tracetree.py
   Example of how to read in a merger tree in HDF5 format and to use a
   specially-defined Tree class to allow spatial and/or temporal searching
   of the tree.
   Usage: python tracetree.py
"""

#--------------------------------------------------------------------------------
# Imports

# The following imports the python3 print function into python2;
# It seems to be quietly ignored in python3
from __future__ import print_function

import sys
import numpy
import h5py

# The tree module contains a class definition for a combined spatial and temporal tree
import tree

#--------------------------------------------------------------------------------
# Parameters
```

```
# The following data set contains a simple example with both a spatial and merger trees.
infile='data/test.hdf5'
# We are going to open read only
mode='r'

#------------------------------------------------------------------------------
# Open Merger tree file for reading
fid = h5py.File(infile, mode)

# This shows how easy it is to extract attributes from the file
for (name,value) in fid['/'].attrs.items(): print(name,value)

# Next we are going to construct a tree from the information contained in the HDF5 file

# First let's extract the number of halos
gid = fid['/MergerTree']
nHalo=gid.attrs.get('NHalo')
print('nHalo =',nHalo)

# and now the links that we want.  Without the [:] we get a pointer to the array.
firstSubhalo=gid.get('FirstSubhaloIndex')[:]
neighbour=gid.get('NeighbourIndex')[:]
firstProgenitor=gid.get('FirstProgenitorIndex')[:]
nextSibling=gid.get('NextSiblingIndex')[:]
hostHalo=gid.get('HostHaloIndex')[:]
descendant=gid.get('DescendantIndex')[:]

# Initialise tree nodes...
# Trick: create an extra non-existent node to receive references to index -1 (= last index)
nodes=[tree.Tree(halo) for halo in range(nHalo+1)]
nodes[nHalo]=None

# Now we run through each of the halos extracting those pointers that we need for
# efficient tree traversal
for halo in range(nHalo):
   nodes[halo].load(nodes[firstSubhalo[halo]],
                    nodes[neighbour[halo]],
                    nodes[firstProgenitor[halo]],
                    nodes[nextSibling[halo]],
                    nodes[hostHalo[halo]],
                    nodes[descendant[halo]])

# The end main halos are those that have no Descendants and no HostHalo
# In this simple example there is only one such halo
endMainHalos=numpy.where((descendant==-1) & (hostHalo==-1))[0]
print(endMainHalos)

# Let's trace all the halos in the combined spatial and temporal tree
nodes[0].printheader()
for halo in nodes[0].next():
   print(nodes[halo])

# Here's a trace of just the spatial tree
nodes[0].printheader()
for halo in nodes[0].next('spatial'):
   print(nodes[halo])
```

```
# and here are the temporal trees for each end halo
endHalos=numpy.where(descendant==-1)[0]
for endHalo in endHalos:
    nodes[endHalo].printheader()
    for halo in nodes[endHalo].next('temporal'):
        print(nodes[halo])

#-------------------------------------------------------------------------------
```

### A4   Tree-search algorithm

We present below an example tree class with methods for generating iterators that can perform a combined (default behaviour), spatial, or temporal tree search.

```
"""tree.py
   Defines a Tree class for dealing with combined temporal and merger trees
"""

#-------------------------------------------------------------------------------
# The following imports the python3 print function into python2;
# It seesm to be quietly ignored in python3
from __future__ import print_function

# In Python 3 the loops of the form
#    for __ in self.something():
#        yield __
# can all be replaced by
#    yield from self.something()
# which makes for much cleaner code.

#-------------------------------------------------------------------------------
class Tree:
    """Tree class for dealing with combined temporal and merger trees"""

    # Constructor: creates an empty tree node.
    def __init__(self,node):
        self.node=node

    # Load pointer arrays.
    def load(self,firstSubhalo,neighbour,firstProgenitor,nextSibling,hostHalo=-1,descendant=-1):
        self.firstSubhalo=firstSubhalo
        self.neighbour=neighbour
        self.firstProgenitor=firstProgenitor
        self.nextSibling=nextSibling
        self.hostHalo=hostHalo
        self.descendant=descendant

    # Default iterator that takes no arguments:
    # Note the asymmetry between spatial and temporal: because there is more
    # than one way to reach a node, can only iterate over one of them.
    # Because of the nature of the trees, space has to be the outer loop.
    def __next__(self):
        yield self.node
        for __ in self.nextSpatial():
            yield __
        for __ in self.nextTemporal('temporal'):
            yield __
```

```python
    # More general iterator method with optional argument specifying tree type:
    def next(self,treeType=None):
        yield self.node
        if treeType is None:
            for __ in self.nextSpatial():
                yield __
            for __ in self.nextTemporal('temporal'):
                yield __
        elif treeType=='spatial':
            for __ in self.nextSpatial('spatial'):
                yield __
        elif treeType=='temporal':
            for __ in self.nextTemporal('temporal'):
                yield __
        else:
            raise valueError('Invalid treeType for tree')
    # Spatial tree search
    def nextSpatial(self,treeType=None):
        if self.firstSubhalo is not None:
            for __ in self.firstSubhalo.next(treeType):
                yield __
        if self.neighbour is not None:
            for __ in self.neighbour.next(treeType):
                yield __
        #raise StopIteration()
    # Temporal tree search
    def nextTemporal(self,treeType=None):
        if self.firstProgenitor is not None:
            for __ in self.firstProgenitor.next(treeType):
                yield __
        if self.nextSibling is not None:
            for __ in self.nextSibling.next(treeType):
                yield __
        #raise StopIteration()

    # Printing
    def printheader(self):
        print ('node,firstSubhalo,neighbour,firstProgenitor,nextSibling,hostHalo,descendant=')
    def __str__(self):
        s1='-1'
        if self.firstSubhalo: s1=str(self.firstSubhalo.node)
        s2='-1'
        if self.neighbour: s2=str(self.neighbour.node)
        s3='-1'
        if self.firstProgenitor: s3=str(self.firstProgenitor.node)
        s4='-1'
        if self.nextSibling: s4=str(self.nextSibling.node)
        s5='-1'
        if self.hostHalo: s5=str(self.hostHalo.node)
        s6='-1'
        if self.descendant: s6=str(self.descendant.node)
        return  str(self.node)+', '+s1+', '+s2+', '+s3+', '+s4+', '+s5+', '+s6

#--------------------------------------------------------------------------------
```

### A5  To extract a tree from an HDF5 file

This example, extracttree.py, reads in an HDF5 file, then extracts some data for a particular tree into a structured numpy array.

```
"""extracttree.py
   Example of how to read in a merger tree file in HDF5 format
   and to create a table containing all the data for a particular tree.
"""

# Note: only if we pick up whole trees will the indices be guaranteed
# to be consecutive.  As we only have a single tree in our test data set,
# this example takes halo 7 to be the root halo as that does have
# consecutive indices for its subtree.

# If the indices are consecutive then there is no need to use a special
# Tree class and methods to allow tree searching.  However, we do first
# do this as a check.

#-------------------------------------------------------------------------------
# Imports

# The following imports the python3 print function into python2;
# It seesm to be quietly ignored in python3
from __future__ import print_function

import sys
import numpy
import h5py

# The tree module contains a class definition for a combined spatial and temporal tree
import tree

#-------------------------------------------------------------------------------
# Parameters

# The following data set contains a simple example with both a spatial and merger trees.
infile='data/test.hdf5'
# We are going to open read only
mode='r'

#-------------------------------------------------------------------------------
# Open Merger tree file for reading
fid = h5py.File(infile, mode)

# This shows how easy it is to extract attributes from the file
for (name,value) in fid['/'].attrs.items(): print(name,value)

# Let's extract the data that we want from the HDF5 file.

# Define numpy data type for the table rows
# Maybe there is a clever way to do this, but this example will just do it by hand
treedata_dtype=numpy.dtype([
        ('haloID','int64'),
        ('descendant','int32'),
        ('firstProgenitor','int32'),
        ('firstSubhalo','int32'),
        ('hostHalo','int32'),
        ('neighbour','int32'),
```

```
            ('nextSibling','int32'),
            ('snapNum','int32'),
            ('density','f8'),
            ('mass','f8')
            ])

# First let's extract the number of halos
gid = fid['/MergerTree']
nHalo=gid.attrs.get('NHalo')
print('nHalo =',nHalo)

# and now the link that we want.  Without the [:] we get a pointer to the array.
haloID=gid.get('HaloID')[:]
descendant=gid.get('DescendantIndex')[:]
firstProgenitor=gid.get('FirstProgenitorIndex')[:]
firstSubhalo=gid.get('FirstSubhaloIndex')[:]
hostHalo=gid.get('HostHaloIndex')[:]
neighbour=gid.get('NeighbourIndex')[:]
nextSibling=gid.get('NextSiblingIndex')[:]
snapNum=gid.get('Snapshot')[:]
density=gid.get('Density')[:]
mass=gid.get('Mass')[:]

# Next we are going to construct a tree from the information contained in the HDF5 file
# This is only needed for trees that are not depth-first ordered, but it's a nice example
# of how to use iterators in python anyway.  It uses the Tree class in the separate tree.py file.

# Initialise tree nodes...
# Trick: create an extra non-existent node to receive references to index -1 (= last index)
nodes=[tree.Tree(halo) for halo in range(nHalo+1)]
nodes[nHalo]=None

# Now we run through each of the halos extracting those pointers that we need for
# efficient tree traversal
for halo in range(nHalo):
    nodes[halo].load(nodes[firstSubhalo[halo]],
                     nodes[neighbour[halo]],
                     nodes[firstProgenitor[halo]],
                     nodes[nextSibling[halo]],
                     nodes[hostHalo[halo]],
                     nodes[descendant[halo]])

# Let's pick a halo to act as the root of our tree.
# Ideally, we should use a full tree, but we only have one tree in this example.
# However halo 7 has consecutive indices in its subtree, so we'll use that.
root=7
nodes[root].printheader()
nTreeHalo=0
nodeMin=nodes[root].node
nodeMax=-1
for halo in nodes[root].next():
    nTreeHalo+=1
    nodeMin=min(halo,nodeMin)
    nodeMax=max(halo,nodeMax)
    print(nodes[halo])
print('nTreeHalo =',nTreeHalo)
print('nodeMin, nodeMax =',nodeMin,nodeMax)
```

```
# Now let's loop over the tree again, recording the data that we want in our table
# Create a numpty structured array to hold the table
treedata=numpy.empty(nTreeHalo,dtype=treedata_dtype)
ihalo=-1
consecutive=True
firstHalo=nodes[root].node
for halo in nodes[root].next():
    ihalo+=1
    if halo != firstHalo+ihalo: consecutive=False
    treedata[ihalo]=(haloID[halo],descendant[halo],firstProgenitor[halo],firstSubhalo[halo],
                     hostHalo[halo],neighbour[halo],nextSibling[halo],
                     snapNum[halo],density[halo],mass[halo])
print('Consecutive halos = ',consecutive)

# If we have a consecutive list of halos, adjust indices to point to the relevant location
# in our new table.
# Setting the maximum to -1 is not really necessary but helps to make things look tidier.
if consecutive:
    numpy.maximum(treedata['descendant']-nodeMin,-1,treedata['descendant']);
    numpy.maximum(treedata['firstProgenitor']-nodeMin,-1,treedata['firstProgenitor']);
    numpy.maximum(treedata['firstSubhalo']-nodeMin,-1,treedata['firstSubhalo']);
    numpy.maximum(treedata['hostHalo']-nodeMin,-1,treedata['hostHalo']);
    numpy.maximum(treedata['neighbour']-nodeMin,-1,treedata['neighbour']);
    numpy.maximum(treedata['nextSibling']-nodeMin,-1,treedata['nextSibling']);

# We have established that we have a consecutive (depth-first ordered) tree,
# so no need to bother with the Tree class again.  Let's just dump the new
# properties in consecutive order as a check.
for ihalo in range(nTreeHalo):
    print(ihalo,treedata[['firstSubhalo','neighbour','firstProgenitor',
                          'nextSibling','hostHalo','descendant']][ihalo])

# Or rather more simply, but in a different order
print(treedata)

#-------------------------------------------------------------------------------
```

### A6  To save a tree in the default format

Finally, we have an example of how to create an HDF5 file in the correct format. Because we need some data to write out, this example also reads in data from an existing HDF5 file: in practice the data may have come from a mergertree creation algorithm.

```
"""dumptree.py
   Example of how to write a merger tree in HDF5 format.
   This example needs some data which it reads in from an existing HDF file,
   which makes this whole exercise a bit pointless, but pretend that the data
   was created some other way and now needs to be written out.
   This is not complete, but it gives the general idea.
"""

#-------------------------------------------------------------------------------
# Imports

# The following imports the python3 print function into python2;
# It seesm to be quietly ignored in python3
from __future__ import print_function
```

```
import sys
import numpy as np

# The merger tree module contains data structures, methods and global variables
import mt

#------------------------------------------------------------------------------
# Parameters

infile='data/test.hdf5'
outfile='data/dumptest_Python.hdf5'

# For the purposes of this example, we first have to read in some data.
# Actually, the following creates pointers to the data, to be read below.

mt_in=mt.MergerTree(infile, "r")
attributes = mt_in.attrs('/')

# Read in snap tables
snap_in=mt_in.read_snaptable()
snapprop_in=mt_in.read_snapprop()
snap_attributes=mt_in.attrs("Snapshots")

# Read merger tree.
tree_in=mt_in.read_mergertree()
tree_attributes=mt_in.attrs("MergerTree")

# Now we are going to create a new HDF5 file to write out the data
print('Creating output file')
mt_out=mt.MergerTree(outfile,"w")
print('File =',mt_out.fid)

# Write out the top-level attributes
print('In group',mt_out.fid.name,'attributes are:')
for (name,value) in attributes:
    print('   ',name,value)
    mt_out.fid.attrs.modify(name,value)

# Create MergerTree directory and set attributes
tree=mt_out.fid.create_group("MergerTree")
print('In group',tree.name,'attributes are:')
for (name,value) in tree_attributes:
    print('name, value =',name,value)
    tree.attrs.modify(name,value)

# Write out merger tree datasets
print('MergerTree datasets:')
for dset in tree_in:
    # extract name, dtype and data from input dataset
    name=tree_in[dset].name
    dtype=tree_in[dset].dtype
    data=tree_in[dset][()]
    attrs=tree_in[dset].attrs.items()
    # and now recreate in the output file
    dset=tree.create_dataset(name=name,dtype=dtype,data=data)
    print('   ',name)
```

```
      print('    ',dset[0:9])
      # and set attributes
      for (name,value) in attrs:
          print('    ',name,value)
          dset.attrs.modify(name,value)

# Create Snapshot directory and set attributes
snap=mt_out.fid.create_group("Snapshots")
print('In group',snap.name,'attributes are:')
for (name,value) in snap_attributes:
    print('    ',name,value)
    snap.attrs.modify(name,value)

# Write out Snap table
# Tables require use of a complex dtype.  Fortuntately, that's already set up
# us here in the input file.
# Extract data and dtype from input file
name=snap_in.name
dtype=snap_in.dtype
nsnap=len(snap_in)
data=np.empty(nsnap,dtype=dtype)
for irow in range(nsnap): data[irow]=snap_in[irow]
# and recreate in output file
dset=tree.create_dataset(name=name,dtype=dtype,data=data)
print('Snap table:')
print('  name =',dset.name)
print('  dtype =',dset.dtype)
print('  data =',dset[:])

# Write out SnapProp table
# us here in the input file.
# Extract data and dtype from input file
name=snapprop_in.name
dtype=snapprop_in.dtype
nsnapprop=len(snapprop_in)
data=np.empty(nsnapprop,dtype=dtype)
for irow in range(nsnapprop): data[irow]=snapprop_in[irow]
# Rename the Units property to lower case (for consistency).
# Cannot rename in place: have to make a copy of both dtype and data.
temp=dtype.__repr__()[6:-1]
temp=temp.replace("Units","units")
dtypenew=np.dtype(eval(temp))
print('dtypenew=',dtypenew)
datanew=np.empty(nsnapprop,dtype=dtypenew)
for irow in range(nsnapprop): datanew[irow]=data[irow]
# and recreate in output file
dset=tree.create_dataset(name=name,dtype=dtypenew,data=datanew)
print('SnapProp table:')
print('  name =',dset.name)
print('  dtype =',dset.dtype)
print('  data =',dset[:])

# The output file is not yet a complete examle of the merger tree data
# format, but hopefuly that is enough to illustrate how do do things.

# Close the files
mt_in.close()
```

```
mt_out.close()

#----------------------------------------------------------------------
```



\label{lastpage}

\end{document}